\documentclass[twocolumn,aps,nopacs,preprintnumbers,nofootinbib, superscriptaddress, amsmath,amssymb]{revtex4}

\usepackage{stmaryrd} 
\usepackage{mathtools} 
\usepackage{stmaryrd} 

\usepackage{amsmath}
\usepackage{amsfonts}

\usepackage{amssymb}
\usepackage{graphicx}%

\begin{document}

\author{ Renan Cabrera}
\email{rcabrera@princeton.edu}
\affiliation{Princeton University, Princeton, NJ 08544, USA} 

\author{Andre Campos}
\affiliation{Princeton University, Princeton, NJ 08544, USA} 
\affiliation{Max Planck Institute for Nuclear Physics, 69117 Heidelberg, Germany}

\author{Herschel A. Rabitz}
\affiliation{Princeton University, Princeton, NJ 08544, USA} 

\author{Denys I. Bondar}
\affiliation{Princeton University, Princeton, NJ 08544, USA} 
\affiliation{Tulane University, New Orleans, LA 70118, USA}

\title{Operational Dynamical Modeling of spin 1/2 relativistic particles: the Dirac equation and its classical limit}

\date{\today}


\begin{abstract}
The formalism of Operational Dynamical Modeling [Phys. Rev. Lett. {\bf 109}, 190403 (2012)] is employed to analyze dynamics of spin half relativistic particles. We arrive at the Dirac equation from specially constructed relativistic Ehrenfest theorems by assuming that the coordinates and momenta do not commute. Forbidding creation of antiparticles and requiring the commutativity of the coordinates and momenta lead to classical Spohn's  equation [Ann. Phys. {\bf 282}, 420 (2000)]. Moreover, Spohn's equation turns out to be the classical Koopman-von Neumann theory underlying the Dirac equation.
\end{abstract}


\maketitle

\section{Introduction}
The Dirac equation is one of the most fundamental building blocks of relativistic quantum theory 
describing the dynamics of spin $1/2$ charged particles. 
The Dirac equation has found a broad range of applications including solid state physics \cite{novoselov2005two,katsnelson2006chiral,hasan2010colloquium}, 
optics \cite{otterbach2009confining,ahrens2015simulation}, cold atoms \cite{vaishnav2008observing,boada2011dirac,anFangzhao-Dirac-BEC}, 
trapped ions \cite{gerritsma2010quantum,blatt2012quantum}, circuit QED \cite{pedernales2013quantum},
and chemistry of heavy elements \cite{Liu-DiracQuantumChem2012,autschbach2012perspective-DiracChem}. 
In this paper we revisit the foundations of relativistic quantum and classical mechanics
to provide a unified operational derivation of the Dirac equation and its classical counterpart, 
addressing the role of spinors and antiparticles in the classical limit $\hbar \to 0$. 

The procedure of applying the limit $\hbar \to 0$ is fraught with many difficulties. 
Considering that $\hbar$ is a fundamental constant with the fixed value, this limit
is a formal procedure whose physical interpretation needs to be clarified. 
The classical limit implies two types of analysis: one involving an equation of motion and the other
-- a quantum state \cite{MBerry2001chaos}. 
The  limit is mathematically ill-defined requiring auxiliary assumptions that may 
significantly change the underlying physical picture \cite{Bialynicki-birula1977}. 
A widely used method to remedy mathematical ambiguities is coarse graining,
which consists of averaging out features of a quantum state arising from interferences.  
This procedure is physically justified by decoherence: erasing quantum coherences by 
coupling the quantum system to an external bath \cite{zurek1991decoherence,jacobs2014quantum}.
However, quantum evolution with decoherence recovers irreversible rather than reversible classical dynamics \cite{cabrera2015efficient}.

Our approach to the classical limit $\hbar \to 0$ of relativistic dynamics is  based on the  
observation that the commutator between the position and momentum of a quantum particle is proportional 
to $\hbar$. This encapsulates 
the Heisenberg uncertainty principle and the experimental fact that
the order of measurements affects the measured outcomes \cite{schwinger2001quantumBook,bondar2011quantumAmJPhys}.
However, the position and momentum of a classical particle can be measured
simultaneously and observed values do not depend on the measurement order.
Mathematically, this implies that the position and momentum of a classical particle commute. Therefore, we define the classical limit as the commutativity of the algebra of observables.
Relativity brings an additional constraint that no antiparticles (i.e., negative energy states) should survive the classical limit. 
This intuition can be formalized by means of Operational Dynamical Modeling (ODM) \cite{BondarODM2012}
-- a universal and systematic framework for deducing physical models from the evolution of
 dynamical average values. 

To derive equations of motion, ODM needs two inputs: observed data recast in the form of Ehrenfest-like relations and kinematics specifying both the algebra of observables and the definition of averages.
As an outcome, ODM guarantees that the resulting
equations have the desired physical structure to reproduce the supplied dynamical observations. 
For example in Ref. \cite{BondarODM2012}, we utilized this method to
infer the Schr\"odinger equation from the Ehrenfest theorems by assuming that the coordinate and momentum
operators obey the canonical commutation relation.
 Otherwise if the coordinate and momentum commute, ODM leads to the Koopman-von Neumann mechanics 
\cite{Koopman1931,Neumann1932,Neumann1932a,mauro2003topics,Gozzi2010,Deotto2003,Deotto2003a},
which is a Hilbert space formulation of non-relativistic classical mechanics where states are represented as 
complex valued wave functions and observables as commuting self-adjoint operators.
ODM has provided a new interpretation of the Wigner function \cite{bondar2012WignerKvN,campos2014violation,flores2016classical}, 
unveiled conceptual inconstancies in finite-dimensional quantum mechanics \cite{bondar2013conceptual}, formulated dynamical models in 
topologically nontrivial spaces \cite{zhdanov2015wigner}, advanced the study of quantum-classical 
hybrids \cite{radonjic2014ehrenfest,gay2018hamiltonian},  quantum speed limit \cite{shanahan2018quantum}, yielded new tools for dissipative quantum 
systems \cite{bondar2016-Dissipation,vuglar2016quantumODM,campos2015excess,zhdanov2016quantumDissipation,zhdanov2017no}, 
and lead to development of efficient numerical 
techniques \cite{cabrera2015efficient,cabrera2016RelativisticWigner,bondar2016efficient,koda2016mixed}.

In the non-relativistic case, ODM relied on the fact that quantum and classical states could be represented on an equal mathematical footing -- the Hilbert space. For the corresponding relativistic program to be carried out,
the state for a spin $1/2$ particle must have a similar representation in the 
quantum and classical realms. Since spinors represent quantum states, the spinorial formulation of classical mechanics is desired \cite{hestenesClassicalSpinor1974,BaylisClassicalSpinor1992,baylisQuantumClassical2010,coddens2015spinors,spohn2000semiclassical,bolte2004zitterbewegung}. 

It is well know that the Dirac equation incorporates spin, but it is uncommon to associate classical dynamics with spin. 
The Lorentz group describes a fundamental symmetry of relativistic mechanics. 
Spinors, also known as ``half vectors,''  are elements of the double cover representation of the Lorentz group \cite{lounesto2001clifford}. 
Classical velocities and accelerations can be expressed in the vector basis formed as bilinear constructions 
of spinors \cite{hestenesClassicalSpinor1974,BaylisClassicalSpinor1992,baylisQuantumClassical2010}. 
Furthermore, there is a specific bilinear combination of these spinors yielding the classical spin, 
whose physical significance is the subject of an ongoing debate \cite{wen2016identifying-classicalSten-Gerlach}.  
Note that there is no spinorial formulation of nonrelativistic classical mechanics except for the Kepler problem \cite{kustaanheimo1965perturbation,hestenes1999Book}. 

This paper is organized as follows: Section \ref{Sec:classical-mechanics}  reviews classical spinorial dynamics. 
Section \ref{Sec:Dirac-Equation} provides an ODM derivation of the Dirac equation. 
Section \ref{Sec:KvND} presents the derivation of the relativistic spinorial Koopman-von Neumann equation. 
Finally, conclusions are drawn in Sec. \ref{Sec:conclusions}.


\section{Classical Mechanics}
\label{Sec:classical-mechanics} 
The purpose of this section is to review relativistic classical mechanics with a particular 
emphasis on the spinorial formulation. The time-extended Lagrangian 
for relativistic classical mechanics 
with electromagnetic interaction  is \cite{fanchi1993review,Greiner1998classicalElectrodynamicsBook,baylis1999electrodynamicsBook}
\begin{align}\label{ExtendedL}
 \mathcal{L} = \frac{m}{2}  u^{\mu}u_{\mu}  + e A^{\mu}u_{\mu} + \frac{1}{2}mc^2,
\end{align} 
where $u^{\mu}$ is the proper velocity, $A^{\mu}$ is the four-vector potential, $m$ is the mass and $c$ 
is the speed of light. 
In this formulation the shell mass $u^{\mu} u_{\mu}=c^2$ is not imposed as a constraint 
but it is instead  incorporated as an integral of motion. The Euler-Lagrange equations
lead to the relativistic Newton equations 
\begin{align}
\label{classical-EulerLagrange}
 \frac{d x^{\mu}}{d\tau} = u^{\mu}, \quad&
m \frac{d u_{\mu}}{d\tau} = e F_{\mu\nu} u^{\nu},  
\end{align}
where $\tau$ is the proper time. The canonical momentum, obtained from 
the Lagrangian is
\begin{align}
 p_{\mu} = m u_{\mu} + e A_{\mu}, 
\end{align}
where we identify $m u^{\mu}$ as the kinetic momentum.  
Note that contravariant indexes are used for physical quantities. 
The time-extended Lagrangian 
can be used to obtain the time-extended classical Hamiltonian $\mathcal{H}$ as
\begin{align}
\label{time-H}
 \mathcal{H} = \frac{1}{2m}  \left(p_\mu - eA_{\mu} \right) 
  \left ( p^{\mu} - eA^{\mu}  \right ) - \frac{1}{2}mc^2.
\end{align}
Assuming no explicit dependence on the proper time, $\mathcal{H}$ 
is a conserved integral of motion corresponding 
to the shell mass condition $\mathcal{H}=0  \longleftrightarrow u^{\mu} u_{\mu} = c^2 $.
The energy $cp^0$ is extracted from the shell mass as
\begin{align}
\label{classical-K}
 c p^0 &= K(p) + c e A_0, 
\end{align}
with the kinetic energy given by 
 \begin{align}
\label{KineticEnergy}
 K(p) = \sqrt{ (mc^2)^2 +  c^2(p  - e A)^k\cdot(p - e A)^k   },
\end{align}
where the Latin indices (e.g., $k$) take values of $1,2,3$.
The Hamilton equations are derived from Eq. (\ref{time-H})
\begin{align}
 \frac{d x^{\mu}}{d\tau}  &= \frac{p^{\mu}-eA^{\mu}}{m}, \\
 \frac{d p_{\mu}}{d\tau} &=  \frac{e}{m} (\partial_{\mu} A_{\nu}) \left( p^{\nu} - e A^{\nu}  \right),
 \label{Hamilton-Eq}
\end{align}
which are equivalent to Eqs. (\ref{classical-EulerLagrange}). 

Classical relativistic mechanics can also be expressed in the spinorial form using two alternative formulations: the
Spacetime Algebra by Hestenes \cite{hestenesClassicalSpinor1974,hestenes1999Book}
and the Algebra of Physical Space by Baylis \cite{BaylisClassicalSpinor1992,BaylisClassicalSpinorEMW199}. 
In this paper we adapt Hestenes' formalism utilizing Feynman's slash notation. The proper velocity is defined as
\begin{align}
   u\!\!\!/ =  u^{\mu}\gamma_{\mu} = u_{\mu} \gamma^{\mu} 
  \label{proper-velocity-u}
\end{align}
where the gamma matrices are $4\times 4$ complex matrices that obey the Clifford algebra in the Minkowski space
\begin{align}
  (\gamma^{\mu}\gamma^{\nu} +\gamma^{\mu}\gamma^{\nu}) = 2g^{\mu \nu} \mathbf{1}
\end{align} 
with $g_{\mu \nu} = diag(1,-1,-1,-1)$. In Feynman's notation, the Lorentz inner product 
is expressed as
\begin{align}
  p^{\mu} q_{\mu} =  p \!\!\!/ \cdot  q \!\!\!/   = \frac{1}{4} Tr[ p \!\!\!/  q\!\!\!/   ],
\end{align}
and the shell mass condition reads
\begin{align}
\label{shell-mass-classical}
 Tr[ (p\!\!\!/ - e A\!\!\!/ - mc  )(p\!\!\!/ - e A\!\!\!/ + mc  )  ] = 0.
\end{align}
A Lorentz transformation of the proper velocity induced by the spinor $L \in \mathbf{Spin}_{+}(1,3)$, 
an element of the double representation of the restricted Lorentz group $SO_{+}(1,3)$
 \cite{lounesto2001clifford}, reads as
\begin{align}
   u\!\!\!/ \rightarrow u^\prime\!\!\!\!\!/ = L u\!\!\!/ L^{-1}.
\label{Lorentz-u}
\end{align}
The spinor $L$ redundantly stores the information. 
In fact, employing the Pauli-Dirac representation of gamma matrices, we have \cite{lounesto2001clifford}
\begin{align}
L = \begin{pmatrix}
    \Psi_{1} & - \Psi_2^{*} & \Psi_3   & \Psi_4^{*} \\
    \Psi_2   &   \Psi_1^{*} & \Psi_4   & -\Psi_3^* \\
    \Psi_3   &   \Psi_4^*  & \Psi_1   & -\Psi_2^* \\
    \Psi_4   &  -\Psi_3^*   & \Psi_2   &  \Psi_1^* 
    \end{pmatrix}, 
\end{align} 
where the column spinor $\Psi$ satisfying the Dirac equation is recovered as
 \begin{align} 
\Psi =   L \Big|_{\text{leftmost column}}. 
\end{align}

It is show in Appendix \ref{AppendixClassSpinor} that
\begin{align}
 \frac{d x^{\mu} }{d \tau}  = u^{\mu}  \Longrightarrow
 \frac{d x^{\mu}}{d\tau} = \Psi^\dagger c \gamma^0\gamma^{\mu}\Psi.
\label{class-spinorial1}
\end{align}
Note that Eq. (\ref{class-spinorial1}) is purely classical even though it resembles relativistic 
Ehrenfest relations. 
The exclusive role of the gamma matrices is to extract the velocity 
stored in the spinor  
\begin{align}
 u^{\mu}  = \Psi^\dagger c \gamma^0\gamma^{\mu}\Psi = 
 \frac{1}{4} Tr(  c L L^\dagger \gamma^0 \gamma^{\mu}).
\end{align}
However, Eq. (\ref{class-spinorial1}) does not imply that the particle is 
moving at the speed of $\pm c$, which are the eigenvalues of $c \gamma^0 \gamma^{\mu}$. The same argument 
holds in the quantum mechanical case, thus eliminating the controversy attributed 
to the use of  $c \gamma^0 \gamma^{\mu}$ as the velocity operator \cite{FockSelected}. 
 
In a similar fashion, the relativistic Newton's equations for the Lorentz 
force in Eq. (\ref{classical-EulerLagrange}) can be recast in the two equivalent forms
\begin{align}
 m \frac{d u_{\mu} }{d\tau}  &=  c e \Psi^{\dagger}  \gamma^0 \gamma^{\nu} F_{\mu \nu} \Psi , \\
 \frac{d p_{\mu} }{d \tau}   &=  c e \Psi^{\dagger}   \gamma^0  (\partial_{\mu}  A \!\!\!/ \,) \Psi. 
\label{class-spinorial2-b}
\end{align}

\section{The Dirac equation}
\label{Sec:Dirac-Equation}
This section offers a derivation of the Dirac equation employing ODM.
According to Ref. \cite{BondarODM2012}, in order to construct
a system's dynamical model, ODM requires the following three inputs: 
\begin{enumerate}
 \item\label{ODMitem1}  The evolution of the average values in the form of Ehrenfest-like relations.
 \item\label{ODMitem2} The definition of the observables' average.
 \item\label{ODMitem3}  The algebra of the observables.
\end{enumerate}

The classical spinorial equations of motion (\ref{class-spinorial2-b}) 
are parametrized in terms of the proper time $\tau$. Considerign that the relation
to the time $t$ is
\begin{align}
 \frac{d}{d \tau} = \gamma \frac{d}{dt},
\end{align}
The classical spinorial equations can be written as 
\begin{align}
\frac{d x^{\mu}}{d t}    = \Psi^\dagger c \gamma^0\gamma^{\mu}\Psi, \qquad
\frac{d p_{\mu} }{d t}   =  c e \Psi^{\dagger}   \gamma^0  (\partial_{\mu}  A \!\!\!/ \,) \Psi,
\end{align}
where the normalization condition $  1 = \Psi^\dagger \Psi$ has been imposed,
resulting in the absorption of the $\gamma$ factor in $\Psi$.
Based on these equations, we \emph{postulate} that relativistic dynamics obeys the following Ehrenfest-like relations:
\begin{align} \label{empirical-Ehrenfest}
 \frac{d}{d t} \left\langle \hat{ \boldsymbol{x}}^{k} \right\rangle = 
  \left\langle   c \gamma^0 \gamma^{k} \right\rangle , \qquad
 \frac{d }{d t} \left\langle  \hat{ \boldsymbol{p}}_{k} \right\rangle 
  =  \left\langle  c e \partial_{k} \hat{A}_{\nu} \gamma^0 \gamma^{\nu}  \right\rangle.
\end{align}
Where $\langle \cdots \rangle$ denotes a physical (empirical) average, which needs to be mathematically defined.  As per item \ref{ODMitem2}, we represent the expectation values by the Dirac bra-ket ``sandwich'' in the Hilbert space,  $\langle \cdots \rangle = \langle \psi | \cdots | \psi \rangle$. Hence,
\begin{align}
\label{rel31-Ehrenfest-2}
 \frac{d}{d t} \langle \psi | \hat{ \boldsymbol{x}}^{k} |\psi \rangle  &= 
  \langle \psi |  c \gamma^0 \gamma^{k}  |\psi \rangle , \\ \label{rel31-Ehrenfest-2b}
 \frac{d }{d t} \langle \psi | \hat{ \boldsymbol{p}}_{k} |\psi \rangle  
  &=  \langle \psi | c e \partial_{k} \hat{A}_{\nu} \gamma^0 \gamma^{\nu}  |\psi \rangle, 
\end{align}
where the position $x^{\mu}$ and momentum $p_{\mu}$ variables are replaced by the corresponding 
 operators $\hat{ \boldsymbol{x}}^{k}$ and $\hat{ \boldsymbol{p}}_{k}$
acting on a spinorial Hilbert space of kets $| \psi \rangle$.

According to the Stone's theorem, unitary evolution of $| \psi \rangle$ implies the existence of 
a self-adjoint  operator $H$ such that
\begin{align}
\label{unitary-evolution-H}
 i \hbar \frac{d | \psi \rangle}{d t}  =  H |\psi\rangle.
\end{align}
Substitution Eq. (\ref{unitary-evolution-H}) into Eqs.  (\ref{rel31-Ehrenfest-2}) and (\ref{rel31-Ehrenfest-2b})
leads to
\begin{align}
  \langle \psi | \frac{1}{i\hbar} [ \hat{ \boldsymbol{x}}^{k}, H] |\psi \rangle &= 
  \langle \psi |  c \gamma^0 \gamma^{k}  |\psi \rangle , \\
  \langle \psi |  \frac{1}{i\hbar} [ \hat{ \boldsymbol{p}}_{k}, H] |\psi \rangle 
  &=  \langle \psi |  ce \partial_{k} \hat{A}_{\nu} \gamma^0 \gamma^{\nu}  |\psi \rangle. 
\end{align}
The expectation values can be dropped assuming that 
these relations are valid for all initial states
\begin{align} \label{CommutatorEq2}
   \frac{1}{i\hbar} [\hat{ \boldsymbol{x}}^{k}, H]  = 
    c \gamma^0 \gamma^{k}  , \quad
    \frac{1}{i\hbar} [ \hat{ \boldsymbol{p}}_{k} , H ]   
  =   c e \partial_{k} \hat{A}_{\nu} \gamma^0 \gamma^{\nu} .
\end{align}
``Quantumness''  is imposed by 
specifying the commutation relations 
\begin{align}
\label{XPCommutator} 
[\hat{ \boldsymbol{x}}^{j} , \hat{ \boldsymbol{p}}_{k} ] = - i \delta^{j}_{\,\,\,\,k} \hbar,
\end{align}
which specifies item \ref{ODMitem3} of ODM.
Note that the negative sign in the right hand side of Eq. (\ref{XPCommutator}) appears because
the nonrelativistic momentum operator is associated with contravariant components $\hat{\boldsymbol{p}}^{j}$
\begin{align}
 {[}  \hat{\boldsymbol{x}}^{k}  ,   \hat{\boldsymbol{p}}_{j}  {]}  = -i \hbar \delta^{k}_{\,\,\, j}  
  \Longleftrightarrow 
 {[}  \hat{\boldsymbol{x}}^{k}  ,   \hat{\boldsymbol{p}}^{j}  {]}  = i \hbar \delta^{k j}.  
\end{align} 
Assuming that $H = H( \hat{\boldsymbol{x}}^k ,  \hat{\boldsymbol{p}}_{k}  ) $,
Eq.  (\ref{CommutatorEq2}) are transformed into the following system of differential equations
\begin{align}
  -\frac{\partial}{\partial \hat{ \boldsymbol{p}}_{k} }  H  = c \gamma^0 \gamma^{k} , \quad
  \frac{\partial}{\partial \hat{ \boldsymbol{x}}^{k}} H =  c e \partial_{k} \hat{A}_{\nu} \gamma^0 \gamma^{\nu}. 
\end{align}
The latter can be readily soled for the unknown generator of motion $H$
\begin{align}
\label{DiracHamiltonian}
 H( \hat{\boldsymbol{x}}^k ,  \hat{\boldsymbol{p}}_{k}  ) 
 = - \gamma^0\gamma^k  c\, \hat{ \boldsymbol{p}}_{k} +   \gamma^0 \gamma^{\nu}  c e  \hat{A}_{\nu} + C,
\end{align}
where $C$ is a constant matrix. 
Note that the obtained $H$ has the dimension of energy. 
Thus, the form of $C$ can be fixed by additionally demanding that the obtained $H$ recovers
the classical Hamiltonian when the position and momentum commutes (i.e., the classical limit).
As shown in the Appendix of Ref. \cite{cabrera2016RelativisticWigner}, this yields
\begin{align}\label{DefCDirac}  
	C = mc^2 \gamma^0.
\end{align}

Finally, note that the equation of motion (\ref{unitary-evolution-H}) with (\ref{DiracHamiltonian}) and (\ref{DefCDirac}) is the sought Dirac equation.

\section{Spin 1/2 Koopman-von Neumann Theory}\label{Sec:KvND}

Having arrived at the Dirac equation, we now find its  classical counterpart.

The classical limit of 
the nonrelativistic quantum state represented by the Wigner function was identified with the Koopman-von 
Neumann wavefunction \cite{bondar2012WignerKvN}. Consequently,  the nonrelativistic classical state 
belongs to a Hilbert space parametrized by both the position and momentum (i.e., the phase space). 
Now we will construct an analog formalism where the classical limit 
of the relativistic Wigner function  corresponds to the \emph{spinorial}
Koopman-von Neumann wavefunction. 

\begin{figure}
  \includegraphics[width=1\hsize]{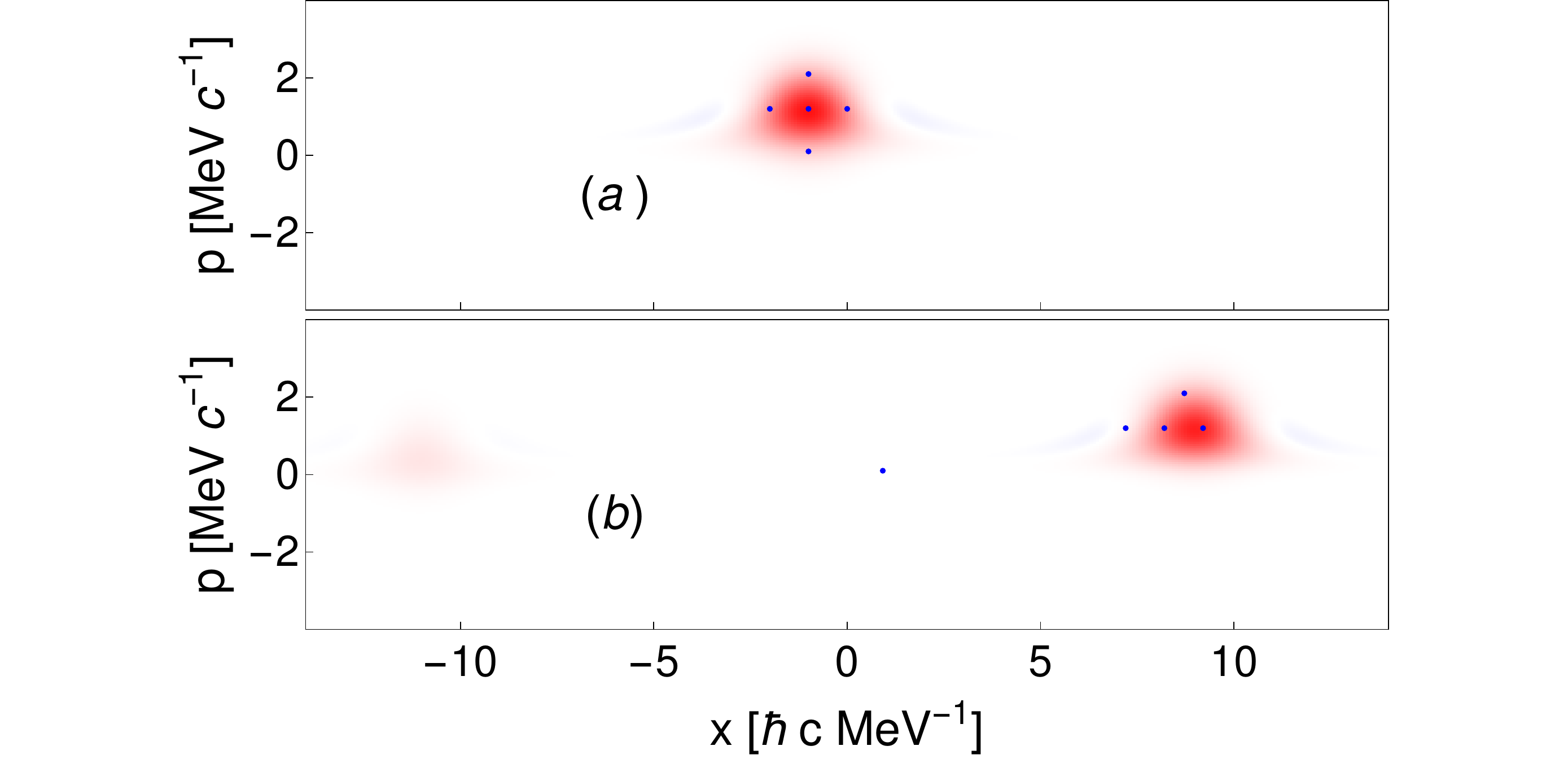}
  \caption{ 
 The Wigner representation of free-particle evolution generated by the (classical) equation of motion (\ref{general-KvN}) and (\ref{DefKnu}) in the phase space. The initial state (a)  obtained by the projecting antiparticles out [Eqs. (\ref{DefW0}) and (\ref{AntiParticleProj})] from a Gaussian state show in Fig. \ref{Fig:DiracFree}(a). The final state (b) contains antiparticles. The blue dots depict an ensemble of point particles evolving according to 
  the Hamiltonian equation (\ref{Hamilton-Eq}). Evolution is restricted to one dimension with $x = x^1$ and $p=p^1$. Red and blue colors represent, respectively, positive and negative values.
 	}
     \label{Fig:FreeKvN-NoFilter}
\end{figure}

\begin{figure}
  \includegraphics[width=1\hsize]{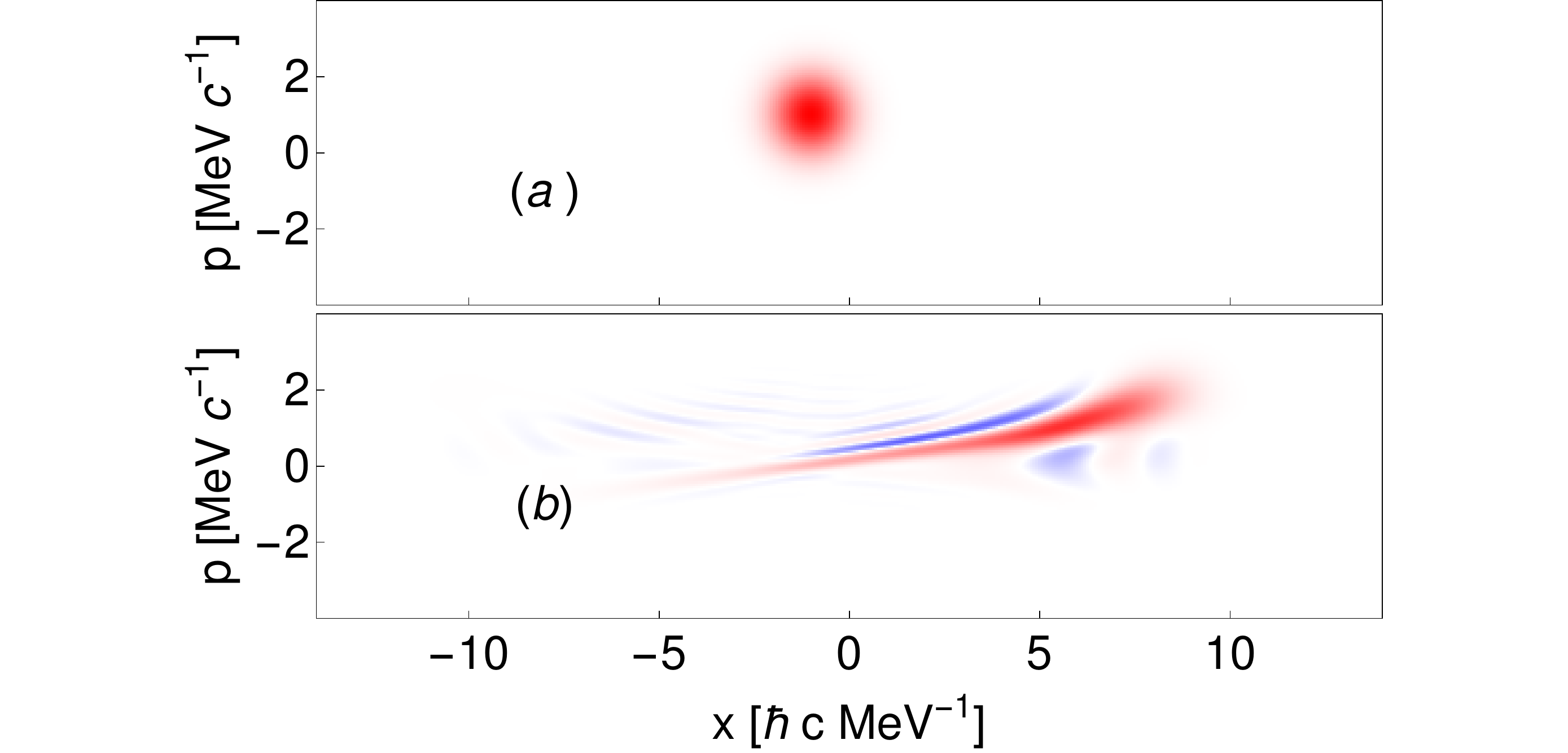}
  \caption{ 
  The Wigner representation of quantum free-particle evolution of the initial Gaussian state (a) via the Dirac equation (\ref{unitary-evolution-H}) and (\ref{DiracHamiltonian}) in the phase space. Evolution is restricted to one dimension with $x = x^1$ and $p=p^1$. Red and blue colors represent, respectively, positive and negative values.
  }
     \label{Fig:DiracFree}
\end{figure}

\begin{figure}
  \includegraphics[width=1\hsize]{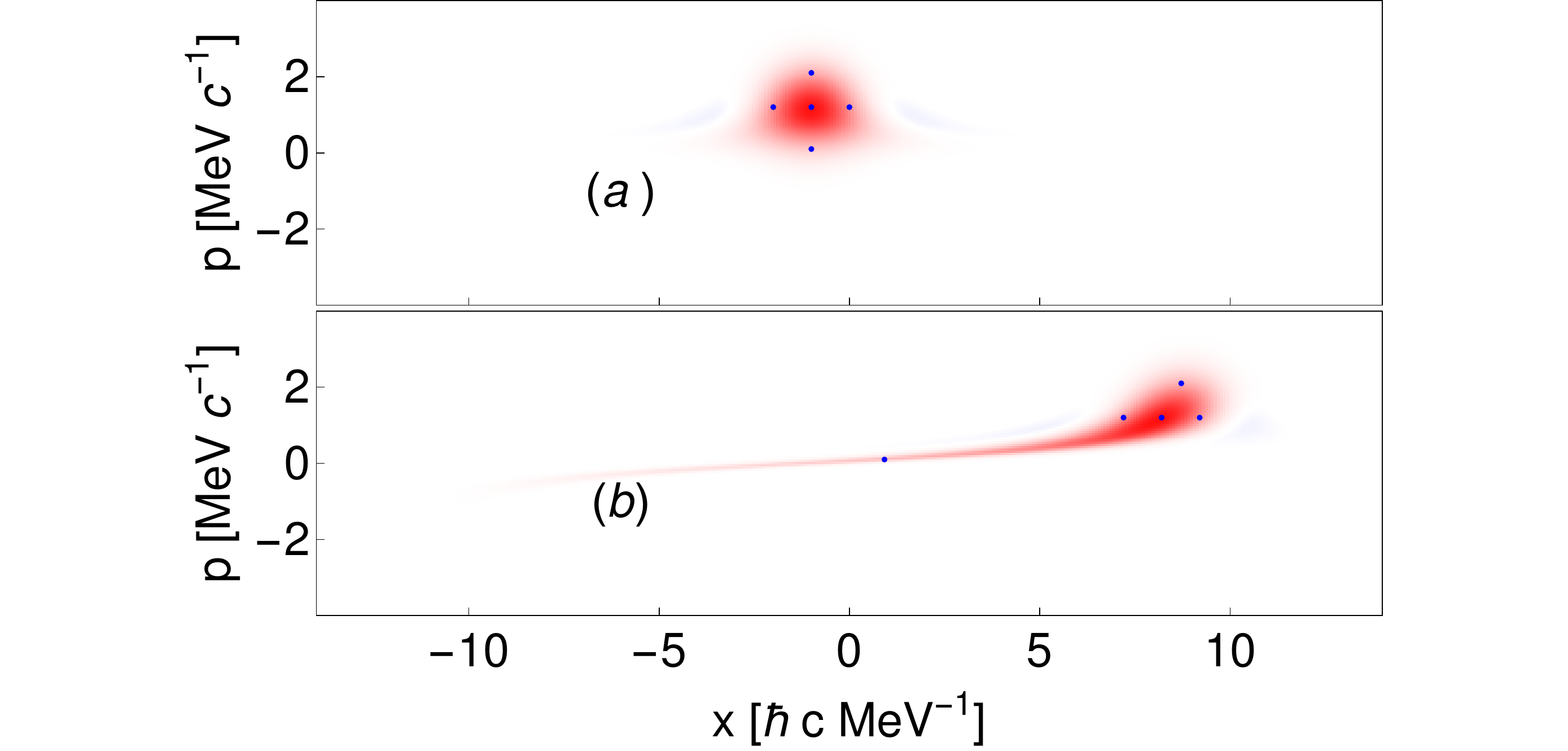}
  \caption{ 
  The Wigner representation of free-particle evolution generated by Spohn's classical equation of motion (\ref{Dirac-KvN-equation}) and (\ref{DefKnu}). The initial state (a)  obtained by the projecting antiparticles out [Eqs. (\ref{DefW0}) and (\ref{AntiParticleProj})] from a Gaussian state show in Fig. \ref{Fig:DiracFree}(a). Unlike the case of Fig. \ref{Fig:FreeKvN-NoFilter}, the final state (b) does not have antiparticles. The blue dots depict an ensemble of point particles evolving according to 
  the Hamiltonian equation (\ref{Hamilton-Eq}). Evolution is restricted to one dimension with $x = x^1$ and $p=p^1$. Red and blue colors represent, respectively, positive and negative values.
  }
     \label{Fig:FreeKvN}
\end{figure}

In this section the physical averages are represented $\langle \cdots \rangle = Tr[  \mathcal{W} \cdots ]$ in terms  of \emph{the Wigner function} $\mathcal{W}$ 
for spin $1/2$  particles,  which is a $4\times 4$ complex matrix \cite{cabrera2016RelativisticWigner,campos2014violation}. 
In this paper, it is convenient to define the Wigner function of a Dirac spinor $\psi(x)$ as
\begin{align}
  \mathcal{W}(x,p) =\frac{1}{2\pi} \int e^{i p \theta} \psi \left(x- \frac{\hbar \theta}{2} \right) 
\psi^{\dagger} \left(x+ \frac{\hbar \theta}{2} \right) d\theta.
\end{align}
\emph{The Wigner representation}, $\mathcal{W}_{\nu,\nu}$ -- the sum of the diagonal elements of the Wigner matrix $\mathcal{W}$, will be used below to visualize dynamics (Figs. \ref{Fig:FreeKvN-NoFilter}, \ref{Fig:DiracFree}, and \ref{Fig:FreeKvN}) since the real-valued function $\mathcal{W}_{\nu,\nu}$ is similar to the non-relativistic Wigner function.

Hence, the Ehrenfest relations (\ref{empirical-Ehrenfest}) read
\begin{align}\label{Ehrenfest-32}
 \frac{d}{d t} Tr[  \mathcal{W} \, \hat{x}^{k} ]  &= 
  Tr[  \mathcal{W}\,  c \gamma^0 \gamma^{k}  ] , \\ \label{Ehrenfest-32b}
 \frac{d }{d t} Tr[  \mathcal{W}\,  \hat{p}_{k} ]  
  &=  Tr[  \mathcal{W}\, c e \partial_{k} A_{\nu} \gamma^0 \gamma^{\nu} ], 
\end{align}
where the trace is calculated over both the spinorial degrees of freedom and the phase space.

Note that in Refs. \cite{cabrera2016RelativisticWigner,campos2014violation} slightly different definitions are used for the Wigner matrix-valued function and representation; additionally,  $Tr$ denotes tracing out the spinorial degrees of freedom only.


``Classicalness'' is introduced by the condition 
\begin{align}
\label{XPCommutator-classical} 
[\hat{ x}^{j} , \hat{ p}_{k} ] = 0.
\end{align}
Similar to the nonrelativistic case \cite{BondarODM2012}, 
the classical algebra must be extended to include additional operators $\hat{\theta}^k$ 
and $ \hat{\lambda}_k$ obeying \cite{cabrera2015efficient,cabrera2016RelativisticWigner}
\begin{align}
  {[}  \hat{x}^{j} ,  \hat{\lambda}_k {]} = -i \delta^{j}_{k} , \quad
  {[}  \hat{p}_{j} ,  \hat{\theta}^k {]} = -i \delta^{k}_{j},
\end{align}
where all the other commutators vanish.

Assuring unitarity of the dynamics, we propose the following anzats for the equation of motion in the classical case
\begin{align}
\label{general-KvN}
 i \frac{\partial }{\partial t} \mathcal{W} = 
 \frac{1}{2}{[}  \gamma^0 \gamma^{\nu} , \hat{K}_{\nu} \mathcal{W}  {]}_{+} , 
\end{align}
where $[ \cdot , \cdot ]_{+}$ is the anticommutator and $\hat{K}_{\nu}$ is an unknown self-adjoint 
generator of motion.
Requiring that in the absence 
of the spinorial degrees of freedom Eq. (\ref{general-KvN}) should reproduce
the non-relativistic Liouvillian equation in terms of the Poisson bracket,
we conclude that $\hat{K}_{\nu}$ must linearly depend on $\hat{\lambda}_{k}$ and $\hat{\theta}^{k}$,  while remaining an arbitrary function of $\hat{x}^k$ and $\hat{p}_k$.
Assuming that $\mathcal{W}$ sufficiently quickly vanishes at infinity, the  generator of motion 
 satisfying the Ehrenfest relations (\ref{Ehrenfest-32}) and (\ref{Ehrenfest-32b}) is
\begin{align}\label{DefKnu}
 \hat{K}_{\nu} = -c \hat{\lambda}_{\nu} - c e (\partial_j A_{\nu}) \hat{\theta}^j,
\end{align}
where $\hat{\lambda}_0=0$.

Even though the obtained model fulfills reasonable conditions, a closer inspection reveals 
that it cannot be a physically valid classical limit. As we will show below, the equation
of motion (\ref{general-KvN}) produces antiparticles.

Antiparticles are convenient to distinguish from particles  in the phase space. Since for the latter, the momentum and velocity vectors are parallel. In other words, a particle with a positive (negative) momentum moves into the positive (negative) direction. However, a portion of the phase space distribution of positive (negative) momenta moving into the negative (positive) direction is associated with antiparticles \cite{cabrera2016RelativisticWigner}. In other words, antiparticle's momentum and velocity vectors are antiparallel, since according to Feynman's characterization, antiparticles are particles moving backwards in time. 

An arbitrary quantum or classical state $\mathcal{W}$ can be made free of all the antiparticle components.  The state
\begin{align}\label{DefW0}
	 \mathcal{W}_0  = \mathcal{P}_{+} \mathcal{W}  \mathcal{P}_{+}
\end{align}
has no antiparticles with the help of the projector \cite{campos2014violation}
\begin{align}\label{AntiParticleProj}
  \mathcal{P}_{+} = \frac{1}{2} \left(
  \boldsymbol{1} + \frac{-\gamma^0\gamma^k c( \hat{p}_k - e A_k)  + mc^2 \gamma^0 }{ K(p)   }  
\right),
\end{align}
where $K(p)$ is the classical kinetic energy defined in Eq. (\ref{classical-K}). For example, the state depicted in Fig. \ref{Fig:FreeKvN-NoFilter}(a), which is the same as Fig. \ref{Fig:FreeKvN}(a), is obtained by projecting a Gaussian shown in Fig. \ref{Fig:DiracFree}(a).

Figure \ref{Fig:FreeKvN-NoFilter}(b) shows a result of free-particle evolution (\ref{general-KvN}) of the initial state  [Fig. \ref{Fig:FreeKvN-NoFilter}(a)] containing no antiparticles. In Fig. \ref{Fig:FreeKvN-NoFilter}(b), one observes two portions of the wave packet containing mostly positive values of momenta but moving into the opposite directions. The left portion consists of antiparticles, whereas particles are on the right. This shows that  Eq. (\ref{general-KvN}) indeed generates antiparticles. As a result, the evolution generated by Eq. (\ref{general-KvN})  disagrees with the classical Hamiltonian evolution (\ref{Hamilton-Eq}) of point particles (see dark blue points in Fig. \ref{Fig:FreeKvN-NoFilter}). 

Dirac free particle dynamics is shown in Fig. \ref{Fig:DiracFree} for comparison. Since the free Dirac evolution does not create antiparticles, all the antiparticles observed in Fig. \ref{Fig:DiracFree} coming from the non-filtered initial Gaussian state in Fig. \ref{Fig:DiracFree}(a).

Nevertheless, the problem of antiparticle creation can be fixed by redefining the Ehrenfest relations as 
\begin{align} \label{Ehrenfest-42}
 \frac{d}{d t} Tr[  \mathcal{W} \, \hat{x}^{k} ]  &= 
  Tr[  \mathcal{W}\,  c \gamma^0 \gamma^{k}  \mathcal{P}_{+} ] , \\ \label{Ehrenfest-42b}
 \frac{d }{d t} Tr[ \mathcal{W}\, \hat{p}_{k} ]  
  &=  Tr[ \mathcal{W}\,  c e \partial_{k} A_{\nu} \gamma^0 \gamma^{\nu}  \mathcal{P}_{+} ].
\end{align}
This leads to the new equation of motion
\begin{align}\label{Dirac-KvN-equation}
 i \frac{\partial }{\partial t} \mathcal{W} = 
 \frac{1}{2} \mathcal{P}_{+}  {[}  \gamma^0 \gamma^{\nu} , \hat{K}_{\nu} \mathcal{W}  {]}_{+} \mathcal{P}_{+}.
\end{align} 
Rewriting the latter as
\begin{align}
 \mathcal{W}(t + \delta t)  = 
  \mathcal{W}(t)
 -i\frac{\delta t}{2} \mathcal{P}_{+}  {[}  \gamma^0 \gamma^{\nu} , \hat{K}_{\nu} \mathcal{W}(t)  {]}_{+} \mathcal{P}_{+} + O\left( \delta t \right),
\end{align} 
we see that if the initial state is free of antiparticles [i.e., $\mathcal{W}(t) = \mathcal{P}_{+} \mathcal{W}(t) \mathcal{P}_{+}$], so is the final state [i.e., $\mathcal{W}(t + \delta t) = \mathcal{P}_{+} \mathcal{W}(t + \delta t) \mathcal{P}_{+}$].
Figure \ref{Fig:FreeKvN} illustrates that the evolution of $Tr(\mathcal{W})$ generated by Eq. (\ref{Dirac-KvN-equation}) does not create antiparticles. An appearance of the tail in Fig. \ref{Fig:FreeKvN}(b) is attributed to the small fraction of the initial wave packet [Fig. \ref{Fig:FreeKvN}(a)] having negative momenta. Furthermore, the wave packet dynamics [Eq. (\ref{Dirac-KvN-equation})] is in agreement with the classical Hamiltonian evolution (\ref{Hamilton-Eq}) of point particles. 

Equation (\ref{Dirac-KvN-equation}) is the classical Koopman-von Neumann theory \cite{BondarODM2012,bondar2012WignerKvN} corresponding to the quantum Dirac equation. The numerical methods for the relativistic Wigner function \cite{cabrera2016RelativisticWigner} (used for Fig. \ref{Fig:DiracFree}) are directly applicable to propagate Eq. (\ref{Dirac-KvN-equation}) as well as Eq. (\ref{general-KvN}).

Equation (\ref{Dirac-KvN-equation}) has been originally derived by Spohn \cite{spohn2000semiclassical} from a different perspective, which established a consistency with the standard classical relativistic mechanics and  the BMT equation for the classical spin \cite{BMT-equation1959}.

\section{Conclusions}\label{Sec:conclusions}

In Refs. \cite{BondarODM2012,bondar2012WignerKvN}, we have reached the conclusion that the value of the commutator between the position and momentum is the only feature distinguishing non-relativistic quantum  form classical mechanics. Here, we have shown that the same conclusion holds in relativistic mechanics. In particular, by starting from the Ehrenfest relations inspired by the spinorial classical mechanics, we deduce the Dirac equation if coordinates and momenta obey the canonical commutation relation. Spohn's  equation \cite{spohn2000semiclassical} is arrived at if in addition to the commutativity of coordinates and momentum (i.e., the classical limit) we explicitly forbid generation of anti-particles. From this point of view, Spohn's  equation emerges as the classical Koopman-von Neumann theory corresponding to the Dirac equation. The develop methodology can be readily apply to the analysis of other relativistic dynamical systems (e.g., governed by the Klein-Gordon equation \cite{kowalski2016wigner,varro2003gauge}).

\acknowledgments

D. I. B. acknowledges a generous supported from AFOSR Young Investigator Research Program (FA9550-16-1-0254).

\appendix

\section{Classical spinor}\label{AppendixClassSpinor}

In classical mechanics, a particle enquires a proper velocity $ u\!\!\!/ $  
by applying a restricted Lorentz transformations  $\bold{Spin}_{+}(1,3)$
on the particle at rest  $ u\!\!\!/_{\text{rest}} = c \gamma^0$
\begin{align}
  u\!\!\!/_{\text{rest}} \rightarrow  u\!\!\!/ = c L L^\dagger \gamma^0, 
 \label{restricted-L}
\end{align}
where $ \frac{d t}{ d \tau} \equiv u^0>0$ implies that the direction of time is preserved.

Any element of $\bold{Spin}_{+}(1,3)$ can be decomposed in terms of a Hermitian ($B$)
and a unitary matrix ($R$)
\begin{align}
 L = BR,
\end{align}
where $B$ is referred to as a Lorentz boost and $R$ as a rotor.
From Eq. (\ref{restricted-L}) we obtain the boost in terms of the proper velocity 
\begin{align}
 B = \sqrt{ \frac{u\!\!\!/ \gamma^0}{c}  }.
\end{align}
The matrix square root can be obtained analytically as
\begin{align}
\label{Boost-B}
 B( u\!\!\!/ ) =
  \frac{ u \!\!\!/ \gamma^0 + \mathbf{1} c (\text{sign} u^0)  }{ \sqrt{ 2 c (\text{sign}u^0) ( (\text{sign}u^0) c + u^0)   }},
\end{align}
where $u^0>0$ for classical particles.

Multiplying Eq. (\ref{restricted-L}) by $\gamma^\nu$ from the right and taking the trace, we obtain 
\begin{align}
 Tr[ \frac{d}{d \tau}x^\mu\gamma_\mu\gamma^\nu ]  &= cTr [ LL^\dagger\gamma^0\gamma^\nu  ] 
 	=cTr [ L^\dagger\gamma^0\gamma^\nu L ] 
 	=cTr [ \gamma^0\gamma^0L^\dagger\gamma^0\gamma^\nu L ] .\label{der1}
\end{align}
It follows that
\begin{align}
  4\frac{d}{d \tau} x^\nu=cTr [ \gamma^0 L^{-1} \gamma^\nu L ],
  \label{der2}
\end{align}
for spinors belonging to the restricted Lorentz transformations. Adding three traceless terms, we have
\begin{align}
   \frac{4}{c}\frac{d}{d \tau}x^\nu = Tr [ \gamma^0 L^{-1} \gamma^\nu L ]
   + Tr [ i\gamma^1\gamma^2  L^{-1} \gamma^\nu L ] \notag\\
   + Tr [  L^{-1} \gamma^\nu L]
   + Tr [ i\gamma^0\gamma^1\gamma^2  L^{-1} \gamma^\nu L ],
\end{align}
Defining the projector  $\mathcal{Q}$ as
\begin{align}
  \mathcal{Q} \equiv \frac{1}{4}(\boldsymbol{1} + \gamma^0 )(\boldsymbol{1} + i \gamma^1\gamma^2 ) =
  diagonal\{1,0,0,0\},  
   \label{projectorQ}
\end{align}
obeying $\gamma^0\mathcal{Q} =  \mathcal{Q}$ and $i\gamma^1\gamma^2\mathcal{Q} =  \mathcal{Q}$,
we arrive to
\begin{align}   
  \frac{d}{d\tau}x^\nu=cTr [ \mathcal{Q} (L^\dagger \gamma^0\gamma^\nu L )\mathcal{Q} ],
   \label{der3}
  \end{align}
which follows from the identity $\mathcal{Q}\mathcal{Q}=\mathcal{Q}$. 

The matrix  $L\mathcal{Q}$ contains $\Psi$ in the first column, while the remainder columns are zero.
Similarly,  $\mathcal{Q}L^\dagger$ contains $\Psi^\dagger$ in the first row, while the remainder rows are zero.
Therefore, Eq. (\ref{der3}) leads to
\begin{align}
   \frac{d x^{\nu}}{d\tau} &= \Psi^\dagger c \gamma^0\gamma^{\nu}\Psi.
     \label{der4}
\end{align}


\bibliography{bib-relativity}

\end{document}